\documentclass[%
 reprint,
 amsmath,amssymb,
 aps,prl
]{revtex4-2}

\usepackage{graphicx}
\usepackage{dcolumn}
\usepackage{bm}

\usepackage[utf8]{inputenc}
\usepackage[T1]{fontenc}
\usepackage{mathptmx}
\usepackage{etoolbox}
\usepackage[english]{babel}

\usepackage{color}

\renewcommand{\selectlanguage}[1]{}
\makeatletter
\def\@email#1#2{%
 \endgroup
 \patchcmd{\titleblock@produce}
  {\frontmatter@RRAPformat}
  {\frontmatter@RRAPformat{\produce@RRAP{*#1\href{mailto:#2}{#2}}}\frontmatter@RRAPformat}
  {}{}
}%

\makeatother
\begin{document}

\preprint{AIP/123-QED}

\title{Active rheology of soft solids performed with acoustical tweezers}
\author{Antoine Penneron}
\homepage{antoine.penneron@u-bordeaux.fr}

\author{Thomas Brunet}%

\author{Diego Baresch}

\homepage{diego.baresch@u-bordeaux.fr}
\affiliation{Univ. Bordeaux, CNRS, Bordeaux INP, I2M, UMR 5295, F-33400, Talence, France} 

\date{\today}

\begin{abstract}
Single-beam acoustical tweezers are used to manipulate individual microbubbles and provide quantitative measurements of the local shear modulus of soft hydrogels. The microbubbles are directly generated by electrolysis of the hydrogel and their displacement is detected using optical  microscopy in the focal plane of a focused vortex beam. Microbubbles displaced off-axis can be pulled by a restoring radial force component that forms a stable two-dimensional trap. We also observe an off-axis tangential microbubble motion that is due to the transfer of the beam's angular momentum flux. A simple elastic model for the hydrogel deformation combined with radiation force calculations finally provide local values of the medium's shear modulus, which are found to be in good agreement with standard bulk measurements performed with a rheometer. Our results suggest that acoustical tweezers are relevant tools to characterize the local mechanical properties of complex soft materials opening new opportunities in the field of active rheology.

\end{abstract}

\maketitle

Local and minimally-invasive measurements of material deformation in response to stress are essential to explore the dynamics of viscoelastic materials. Indeed, a wide range of soft materials including foams, suspensions or biological tissues exhibit mechanical properties that depend on the length scale at which they are observed, and cannot be fully characterized with standard macroscopic bulk measurements \cite{squires2010fluid,waigh2016advances}.  

Active rheology measurements, where an external force is applied to a small probe, are well adapted for the local  micromechanical characterization of complex soft materials -- \textit{e.g.} heterogeneous, anisotropic or fragile-- in situations where the direct access to the sample is difficult or impossible. Optical and magnetic tweezers are standard implementations that can apply a contactless force on particles of size lying between 100~nm  and 10~$\mu$m \cite{neuman2008single,furst2017microrheology}. Optical tweezers are very precise and versatile tools, but are limited to forces of a few hundreds of piconewtons and require the use of high intensity laser beams. Magnetic tweezers slightly increase the force range available for micromanipulation tasks, but they are usually limited to the application of a unidirectional pulling force and suffer from a poor spatial resolution in most conventional setups. In both cases, a direct optical access to the probe's position is required, limiting their use to sufficiently transparent samples or to measurements performed near their surface.

The radiation force generated by sound waves is in comparison much larger and has been used for decades in a diversity of applications \cite{marston2004manipulation, bruus2011forthcoming, thomas_acoustical_2017, dholakia2020comparing}. Ultrasonic beams can be focused deep into absorbing tissue and directly apply a remote `palpation' force on macrsocopic scale volumes ($\sim$mm$^3$) useful for clinical characterization of tissues \cite{nightingale2011acoustic,gennisson2013ultrasound}. Solid spherical probes have directly been embedded in deformable materials and pushed in the propagation direction of the beam, providing local, yet macroscopic, measurements of rheological properties of viscoelastic materials \cite{aglyamov_motion_2007, urban2011generalized,yoon2011, suomi2015optical}. The complex rheological behavior of yield-stress fluids has also been recently explored using acoustically forced solid probes \cite{lidon_measurement_2017} or bubbles \cite{saint2020acoustic}, and highlight the relevance of ultrasound to assess the (far) out-of-equilibrium rheological response of a material over previously inaccessible length scales \cite{lidon2019mesoscale}. In some of the previous examples, the acoustic forcing has been integrated with an ultrasound-based echolocation strategy to track the probe's position with a good spatio-temporal accuracy, paving the way to active rheology measurements in opaque-to-light materials. However, it has just been over a decade that it was demonstrated that focused ultrasonic beams could be utilized to form acoustical tweezers for precise three-dimensional particle  manipulation in simple liquids \cite{baresch_observation_2016, thomas_acoustical_2017}, with a working principle similar to the single-beam gradient force optical trap, also know as optical tweezers \cite{ashkin_observation_1986}. Despite several interesting advantages, the suitability to perform quantitative local rheology measurements with acoustical tweezers has not yet been explored. 

In this Letter, we describe the use of single-beam acoustical tweezers to drive the motion of individual microbubbles used as probes typically 100 $\mu$m in size embedded in a hydrogel. We apply radiation forces in the sub-micronewton range and optically detect microbubble displacements of a few microns. A vortex beam is used to form a two-dimensional trap in the focal plane by attracting the microbubbles towards the propagation axis with a restoring radial force. For bubbles positioned off-axis, we show that an azimuthal force component of similar magnitude is applied and drives a tangential motion. We use radiation force calculations to evaluate the force that displaces the microbubbles. Combined with a model for motion of the microbubble in an elastic medium leads to quantitative measurements of the local shear modulus of several hydrogels. The local measurements are in good agreement with bulk values obtained independently with standard shear rheometry. By combining the main advantages of optical trapping with the appealing characteristics of ultrasound propagation, the present work paves the way for an acoustic active rheology approach that offers interesting perspectives for the characterization of a wide range of complex soft materials.   


The motion equations to estimate the static and transient displacement of a solid sphere or gas bubble embedded in a viscoelastic medium in response to an applied force were derived by Ilinskii and co-workers \cite{ilinskii_gas_2005}. For negligible bubble surface deformation, the applied radiation force, $\mathbf{F}$, and the displacement of the bubble's center, $\mathbf{u}$, can be related as:         
\begin{equation}
\mathbf{F} = 4 \pi G a \mathbf{u},
\label{equation:Stokes-Types}
\end{equation}
where $a$ is the bubble radius and $G$ the medium's shear modulus. Eq.\ref{equation:Stokes-Types} has been used to calibrate the applied force in soft gels by measuring particle displacements with a prior knowledge of $G$ \cite{lidon_measurement_2017}, and, reversely, to measure the local value of $G$ provided the applied force is known \cite{saint2020acoustic}. This latter approach is adopted in what follows.

\paragraph*{Experimental setup and methods}

\begin{figure}
\includegraphics[width=1.0\linewidth]{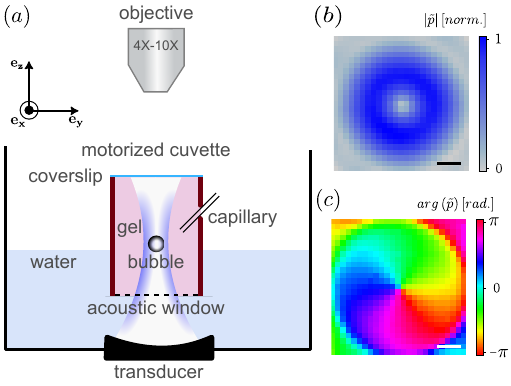}
\caption{ a) Schematic of the experimental setup (not to scale). A mobile cuvette containing the hydrogel is mounted on motorized translation stages and immersed in a water tank. A vortex beam is generated by a concave array of 8 piezoelectric transducers and focused through an acoustically transparent window into the gel. The microbubble is generated by electrolysis of the hydrogel at the tip of an electrode inserted through a small lateral capillary. The same capillary can be used to insert a fibre optic hydrophone. The microbubble position is tracked using a long working distance optical objective and a camera. b)-c) Experimental magnitude, $|\tilde{p}|$, and phase maps, $arg(\tilde{p})$, of the incident vortex beam ($\ell=+1$, $f=2.25$ MHz) obtained by scanning the hydrophone in the focal plane. The scan step is set to $75~\mu$m. The scale bars represent $300~\mu$m.}
\label{fig_1}
\end{figure}

The acoustical trap is formed by directly focusing a vortex beam \cite{baresch_observation_2016} into a small 3D-printed cylindrical cuvette (diameter: 30~mm, height: 30~mm) filled with a hydrogel (Fig.\ref{fig_1}a). It is equipped on its bottom with an acoustically transparent window made of a thin polyethylene film (thickness: $\sim 10~\mu$m) and sealed by a glass coverslip on its top that ensures a clean and smooth optical imaging interface. The hydrogel is prepared by dissolving a polymer (Carbopol ETD-2050) in distilled water at low mass concentrations (0.3 to 1$\%$ wt) following a protocol described elsewhere \cite{geraud2013confined}. The gel subsequently centrifuged to remove any residual bubble, left to stabilize for several days and finally poured in the cuvette before starting the experiments. The selected polymer concentrations in this work result in a yield-stress fluid capable of entrapping bubbles up to a few mm in size against gravity, while maintaining acoustic propagation properties very similar to those of water (propagation speed $c\sim$ 1500~m/s at room temperature), with a low attenuation at MHz frequencies ($\alpha \sim 5\times 10^{-5}$ MHz$^{-2}$mm$^{-1}$ \cite{brunet2012sharp,tallon2020energy}). Such a low attenuation limits potential absorption driven heating effects or radiation force deformation of the gel itself. The elastic shear modulus of the hydrogels is characterized performing oscillatory bulk measurements using a rotational shear rheometer. More information on the measurements and data can be found in the Supplementary Material.

The focused vortex beam is generated by an array of 8 piezoelectric transducers (Imasonic, France) arranged as radial sectors on a concave surface (diameter: 40~mm, radius of curvature: 30~mm, $F$-number: 0.75) located at the bottom of a water tank (width: 20~cm, length: 20~cm, height: 15~cm), and operated at its central frequency $f = 2.25$~MHz (wavelength: $\lambda=c/f=660~\mu$m) with adequate phase shifts to generate a beam of topological charge $\ell=+1$ \cite{baresch_acoustic_2020} using an arbitrary signal generator (Image Guided Therapy, France). The pressure amplitude, $p$, can be set between 50 and 150~kPa, a range for which the propagation remains linear. We also checked their was no detectable distortion or attenuation of the wavefront in our hydrogels compared to propagation in water. The cuvette is equipped with a lateral plastic capillary (diameter: $\sim 1$~mm) by which an optical fiber hydrophone (diameter: 100 $\mu$m, sensitive element size: 10 $\mu$m, Precision Acoustics, UK) is inserted to measure the pressure profile by moving the cuvette relative to the fixed acoustic focus with motorized translation stages. The acoustic field can be described by analyzing the magnitude and phase of the Fourier transform, $\tilde{p}$, applied to the received short duration bursts (10 cycles) at the central frequency. The normalized magnitude profile shows the typical high intensity ring of a vortex beam (Fig.\ref{fig_1}b) surrounding the core where the intensity drops to zero due to destructive interferences near the phase singularity (Fig.\ref{fig_1}b). Note also that straight radial lines of constant phase can be drawn from the singularity outward and are a useful characteristic of the vortex field in the focal plane. Therefore, during this step, we colocalize the acoustic focal plane with the optical imaging plane of a high working distance, low depth-of-field objective (magnification: 4X-10X, OPTEM Fusion) with a good precision ($<75~ \mu$m $\sim \lambda/8$) by recording the position of the hydrophone tip with a camera (PCO Panda 4.0). The water tank is illuminated by a LED panel that provides a homogeneous imaging background.

We use a simple setup consisting of a pair of copper wires (diameter: 144~$\mu$m) and a 9-volt battery to directly generate a microbubble by electrolysis of the hydrogel. A first wire is inserted through the capillary and guided towards the cuvette center where acoustic interactions between the bubble and the walls are minimized, while the second wire is allowed to rest in the water-tank. Microbubbles with a typical radius in the range $a \sim 50 - 100~\mu$m are obtained by the successive coalescence of smaller bubbles at the tip of the wire. The wire is then quickly removed to leave a single microbubble near the center of the cuvette. Note that all the microbubbles are driven well above their natural resonance frequency that is less than 100~kHz for the size range and hydrogel elasticities we consider \cite{dollet2019bubble}.  We additionally observe that it is necessary to switch the electric polarity during the bubble generation process in order to efficiently obtain large bubbles. We therefore do not control the final mixture of oxygen and hydrogen, but we checked its  effect on the resulting radiation force (See Supplemental Material for more information). Finally, the initial bubble position in the acoustic focal plane can be fine-tuned using the motorized translation stages with an accuracy of a few microns.

To process the optical images, we first apply a segmentation algorithm (Ilastik \cite{berg2019ilastik}) to accurately isolate the microbubble from the image background. We then retrieve the bubble centroid, $\mathbf{x}$, and mean radius, $a$, using open access image analysis tools (Fiji \cite{schindelin2012fiji}). Overall, we obtain a good sub-pixel centroid localization precision of approximately 0.2 $\mu$m thanks to the relatively large microbubble size and good spatial resolution of the image (0.9 $\mu$m per pixel). To displace the microbubble, we apply a much longer ultrasonic burst continuously for a duration of 0.5~s that is sufficient to ensure the final equilibrium position is reached. That is, when the hydrogel generates an elastic resistance force that exactly offsets the applied radiation force. The displacement is computed as $\mathbf{u(\mathbf{r_0})} = \mathbf{r} - \mathbf{r_0}$ where $\mathbf{r}$ and $\mathbf{r_0}$ are the centroid positions during ultrasound application and initial position within the acoustic field respectively. We also implement a test of the small elastic deformation hypothesis by analyzing the residual displacement $\mathbf{u'(\mathbf{r_0})} = \mathbf{r_0} - \mathbf{r_0'}$ where $\mathbf{r_0'}$ is the centroid position obtained from an image taken after the acoustic excitation is turned off and the bubble allowed to relax to a new position for a duration that exceeds the hydrogel's relaxation time. For the different hydrogel concentrations, the driving pressure is set to remain in the elastic regime, and we discard any experiment for which the magnitude of $\mathbf{u'}$ is larger than our typical position detection noise ($\sim$ 0.2~$\mu$m, Supplementary Figure~4).    

To compute the force applied on the microbubbles, we implement a model based on a general three-dimensional (3D) framework to compute the radiation stress exerted on a spherical particle arbitrarily located in a compressible incident field \cite{baresch_three-dimensional_2013}, but extended to include the effect of the surrounding soft elastic medium on the scattered compressional wave \cite{ilinskii_acoustic_2018}. For bubbles of arbitrary size relative to the wavelength, the model yields the vector components of the force applied in the Cartesian coordinate system centered on the bubble as a function of its position in the focused beam. More details on the model and its implementation can be found in the Supplementary Material.

\paragraph*{Results}

\begin{figure*}
\includegraphics[width=1.0\linewidth]{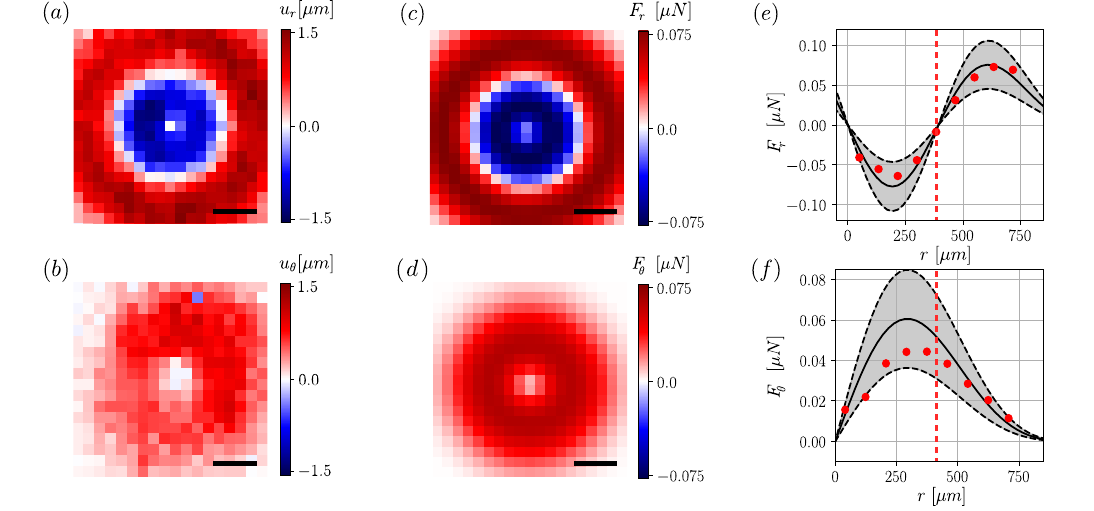}
\caption{(a)-(b) Radial $u_r$ and azimuthal $u_\theta$ displacement maps obtained in the focal plane $(X,Y)$  for a microbubble of mean size $a=80 \mu$m and an incident acoustic pressure $p=110$~kPa} measured at the vortex ring shown in Fig.~\ref{fig_1}(b). The measurement step is set to $75 \mu$m. (c)-(d) Computations of the theoretical radial $F_r$ and azimuthal $F_\theta$ force components using the experimental values of $a$ and $p_{ac}$. (e)-(f) Experimental value for $F_r$ and $F_\theta$ (circles) obtained using Eqn.\ref{equation:Stokes-Types} and the best-fit value of $G$ that reproduces the  modeled radial force component. The values are obtained at each radial position of the bubble centroid after an average over all $\theta $ directions. Confidence boundaries on the fit are set by the pressure measurement uncertainty of $17 \%$ resulting in a $34\%$ uncertainty on the force. (f) Overlay of the experimental and theoretical values of $F_\theta$ using the fitted value for $G$. The dashed vertical red line in (e) and (f) indicates the radial position of the vortex ring of maximum intensity. The scale bars represent $300~\mu$m.
\label{fig_2}
\end{figure*}

We first show in Figure \ref{fig_2}, for a typical experiment, the microbubble displacement as its initial position is scanned in the focal plane. The displacement is decomposed into a map for the radial component, $u_r$, and for the azimuthal component, $u_\theta$, defined in a polar coordinate system $(r,\theta)$ originating at the vortex core in the focal plane as shown in Fig.\ref{fig_2}~(a). Such a decomposition is not surprising in vortex beams since, on one side the specific intensity distribution shown in Fig.\ref{fig_1}~b) generates a force component in the direction of the radial intensity gradient \cite{baresch_observation_2016,baresch_acoustic_2020} and, on the other, the beam's angular momentum flux generates an azimuthal scattering force component \cite{baresch_spherical_2013}. Our observations clearly confirm this behavior for microbubbles embedded in an elastic medium. Fig.\ref{fig_2}~(a) additionally shows that the bubble is attracted towards the vortex core up to a distance that is equal to the radius of the maximum intensity ring, forming a stable 2D trap, and repelled beyond. This is consistent with the out-of-phase response of a bubble driven above its natural monopolar resonance frequency that leads to a force in the direction of local pressure nodes (see \textit{e.g.} Refs. \cite{asaki1994,saint2020acoustic}). The azimuthal displacement component (Fig.\ref{fig_2}b) is exclusively positive and maximum when the bubble is positioned close to the vortex ring radius, where the azimuthal component of the average energy flux density, $\left<S_\theta\right>=\left<p v_\theta\right>$, where $v_\theta$ is the azimuthal component of the particle velocity field, has a high amplitude. Force calculations (not shown) indicate that the axial component $F_z$ does not form a stable trap in the propagation, but can push the microbubble instead. The induced out-of-plane motion was however not sufficient to be readily observed or affect our in-plane motion analysis.  Close to the propagation axis, $F_z$ is predicted to be much weaker because the monopolar bubble dynamics cannot be efficiently forced in vortex beams \cite{baresch_acoustic_2020}.

To test the linear dependence between bubble displacement and applied force suggested by Eq.(\ref{equation:Stokes-Types}), we now compute the theoretical radiation force using experimental values of $a$ and $p$. In principle, the model also requires the prior knowledge of the soft medium's shear modulus at high frequencies ($\sim$MHz) that theoretically can have an impact on the microubble oscillation dynamics and resulting force. These unconventional values are not available for Carbopol, but we checked that for sufficiently soft hydrogels $G< 1$~MPa, there was no appreciable effect on the force for a wide range of bubble sizes, in agreement with recent experimental observations for bubbles in the long wavelength scattering regime \cite{saint2020acoustic}.   

Figs. \ref{fig_2}(c) and (d) show plots of the computed radial, $F_r$, and azimuthal, $F_\theta$, components obtained at each bubble position scanned in experiment. Both components highlight the good agreement between the force predictions and measured bubble displacements and confirm their linear relation. The shear modulus $G$ of the surrounding medium is the only free parameter, and we extract its local value $G=50 \pm 20 $~Pa with a least-square fit of our experimental data for $F_r$ following Eq.(\ref{equation:Stokes-Types}). This value is in good agreement with that obtained from standard bulk oscillatory rheology, $G=43$~Pa (Supplementary Material Fig.~3). The fit quality can be better appreciated in Fig. \ref{equation:Stokes-Types}(e), where we overlay the predicted value for $F_r$ with our measurements as a function of the radial position of the bubble, and averaged over all $\theta$ directions. Additionally, the fitted value for $G$ also reproduces accurately the expected value for $F_\theta$ in Fig.~\ref{fig_2}(f), increasing further the confidence on the fit goodness. For both components, the dashed lines show the force curves obtained by applying an uncertainty of $\pm 17\%$ to the measured pressure value $p=110$~kPa and used for the computation. The confidence boundaries for the estimation of $G$ are therefore set by an uncertainty of $\pm 34\%$ as the applied force scales with $p^2$. The uncertainties related to the measurement of $a$ and $\mathbf{u}$ are negligible in comparison.

\begin{figure}
\includegraphics[width=1.0\linewidth]{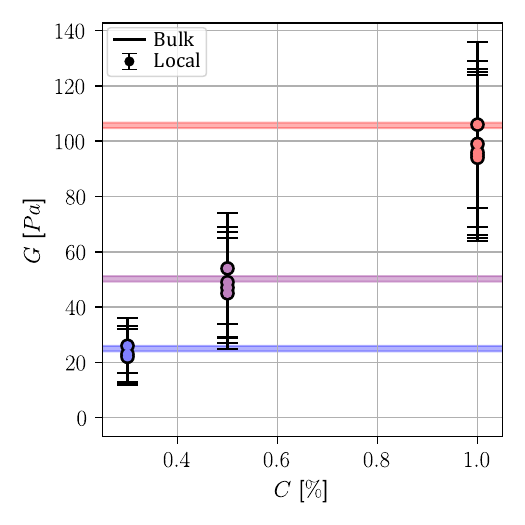}
\caption{Estimation of the local shear modulus $G$ of hydrogels prepared using three different concentrations: $C=0.3$, 0.5 and 1.0$\%$. Circles represent the data obtained from the fit of the experimental radial force component with the model. The uncertainty is set by the 34\% error on the applied force, the size of the bubbles $a$, and magnitude of the displacement following Eqn.\ref{equation:Stokes-Types}. The horizontal lines correspond to the low-frequency limit of the shear modulus obtained with oscillatory shear rheometry (Supplementary Material Fig.~3).}
\label{fig_3}
\end{figure}

We report in Fig.~\ref{fig_3} estimations for the local value of $G$ in hydrogels made using three different carbomer concentrations $C=0.3, 0.5$ and $1.0\%$~wt. For all experiments the radiation force is kept low enough to remain in the elastic deformation regime by setting the acoustic pressure to respectively $85$, $110$ and $130$~kPa, yet sufficient to induce a displacement magnitude that exceeds the localization precision threshold. By exploiting the radial displacement component as previously described, we obtain respectively $G= 24$, 49 and 98~Pa for the local modulus averaged over independent experiments performed with respectively $n=3$, 4 and 5 bubbles. These values are all in good quantitative agreement with the low-frequency limit of those obtained from oscillatory bulk rheology measurements (Supplementary Material Fig.~3). The uncertainty estimates originates again from the acoustic pressure measurements and tends to increase for higher forcing amplitudes. Nevertheless, the data shown suggest a good reproducibility of our results for the three gels and for all bubble sizes explored.    

\paragraph*{Discussion and conclusion}
The results presented in this Letter show that acoustical tweezers relying on a focused propagating vortex beam are a relevant tool for precise and local measurements of the mechanical properties of soft materials. The local shear modulus of a range of hydrogels with varying elastic properties was successfully measured and compared to bulk values in the elastic deformation regime. 

The technique is based on the application of a controlled force on a probe particle with subsequent tracking of its displacement, thus belonging to the class of active rheology measurements. Compared to standard methods with either optical or magnetic tweezers  \cite{neuman2008single,waigh2016advances}, acoustical tweezers present several appealing advantages for soft material characterization. Ultrasonic beams have a large penetration depth ($>10$~mm) in many opaque-to-light media while retaining a relatively good resolution ($10-1000~\mu$m) at common transducer frequencies ($1-100$~MHz). They can also apply high magnitude forces at moderate field intensities, provided the probe particle has a good acoustic contrast with the propagation medium and that intensity gradients resulting from a high degree of focusing are steep \cite{thomas_acoustical_2017,dholakia2020comparing}. The contrast arises from compressibility and density differences between the probe and the propagation medium, which are physical properties that are easy to control (compared \textit{e.g.} to magnetization) and could extend the variety of probes that can be used. The exceptional contrast of microbubbles in the medium allowed to exert forces in the submicronewton range  with acoustic intensities remaining below 1 W/cm$^2$, thus avoiding absorption-driven heating effects that could affect the measurement. Compared to the very high intensities ($\sim 10^9-10^{12}$~W/cm$^2$ \cite{neuman2008single}) required at the focus of an optical trap, acoustic tweezers arguably  highly limit the risk of medium heating and damage. Nevertheless, the exact mechanisms leading to sample damage with an acoustic trapping beam remain to be thoroughly investigated as they result from very different physical processes. Importantly, the lower contrast of a solid probe particle such as \textit{e.g.} polystyrene would result in a force nearly 4-orders of magnitude weaker compared to a microbubble. Additionally, the reduction of the probe size results in a drastic decrease as the gradient force component scales with the particle volume ($\sim (a/\lambda)^3$) in the sub-wavelength scattering regime. Therefore, future investigations should evaluate the effectiveness of the approach for accurate measurements at the microscale when using smaller probe particles ($<10~\mu$m) and higher frequency ultrasonic beams ($>10$~MHz) with a careful attention paid to the resulting force, beam attenuation, and potential heating of the probed material. 

 A prior knowledge of the applied force is essential to accurately estimate the local mechanical properties of the material. This was managed by using spatially resolved acoustic pressure measurements combined with a radiation force model \cite{baresch_three-dimensional_2013,baresch_spherical_2013,ilinskii_acoustic_2018}. The well-controlled 2D trapping behavior we reported, and particularly the expected linear variation of the radial force with small off-axis bubble displacements, could additionally serve as a basis to test alternative force calibration approaches and narrow down measurement uncertainties. 
 
 An interesting perspective of this work is the coupling of the acoustical trap with ultrasonic imaging techniques to provide an "all acoustical" active rheology measurement in opaque-to-light materials. Other future research directions include using higher frequency vortex beams for manipulation at the microscale \cite{baudoin_spatially_2020}, exploring trap calibration strategies, probing nonlinear mechanical responses and developing shear rheology approaches with spinning probes by exploiting the transfer of angular momentum \cite{bishop2004optical,anhauser2012acoustic,baresch2018orbital, alhaitz2023confined}.

\paragraph*{Acknowledgments}
We are grateful to T. Salez, J-L. Thomas, Y. Amarouchene P. Lidon and G. Ovarlez for fruitful discussions. We also express our gratitude to E. Laurichesse for his assistance in obtaining the bulk rheology of our samples, and E. Dumont from \textit{Image Guided Therapy} for providing the multi-channel generator. D.B. is supported by the Agence Nationale de la Recherche (ANR-22-CE30-0031). This article is submitted under the Creative Commons Attribution-NonCommercial-ShareAlike 4.0 International License ( https://creativecommons.org/licenses/by-nc-sa/4.0/).

\bibliography{APL}
\clearpage

\onecolumngrid

\section{Supplementary Material}
%
%
%
%
%
%
%

\maketitle

\section{Radiation force calculation using experimental acoustic pressure measurements}
To compute the acoustic force exerted on the bubble arbitrarily positioned in the incident vortex beam, we used the method developed in Refs. \cite{baresch_three-dimensional_2013,baresch_spherical_2013}. This model was established for sound waves propagating in simple fluids. Ilinskii \textit{et al} recently considered the more general case of the force generated in an elastic solid by an incident compressional wave \cite{ilinskii_acoustic_2018}. Nevertheless, we find that for the very soft hydrogels we considered ($G<100$~Pa), the effect of the surrounding elasticity is negligible as far as the sole role of compressional waves is considered to compute the force. Conversion to shear waves during the scattering process should however be important for stiffer propagation media. The interested reader can refer to Ref.\cite{ilinskii_acoustic_2018} for useful discussions on the limitations of the model.            

\begin{figure}
\centering
\includegraphics[width=0.6\linewidth]{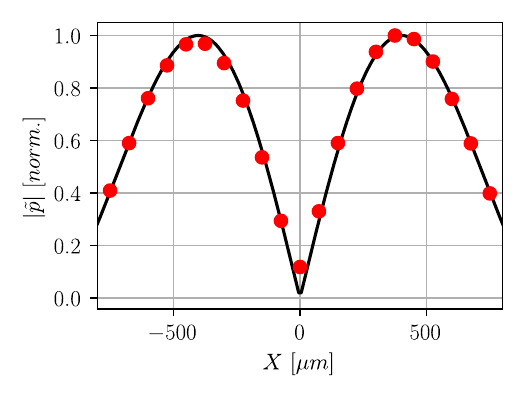}
\caption{Comparison of the modeled pressure magnitude (plain line) with hydrophone measurements (red dots) obtained for the incident vortex beam ($\ell=1, f=2.25$~MHz) in the focal plane. The hydrogel concentration, $C$, has no observable effect on the beam propagation characteristics for the soft gels we used ($G<100$~ Pa). }
\label{fig_supp_4}
\end{figure}

In brief, the monochromatic incident beam is considered to propagate linearly and is expanded in a set of spherical harmonics in the spherical basis centered on the bubble $(r, \theta,\varphi)$ as:
\begin{equation}
\phi  = \phi_0 e^{-i\omega t}\sum_{n = 0}^{\infty}\sum_{m = -n}^{n}A_n^m j_n(kr)Y_n^m(\theta,\varphi),
\label{eq:phi}
\end{equation}
where $\phi$ is the complex acoustic velocity potential related to the incident particle velocity and pressure as $\mathbf{v}=\mathbf{\nabla\phi}$  and $p=-\rho\left( \frac{\partial \phi}{\partial t}\right)$ respectively, and the time convention, $e^{-i\omega t}$, is used. $\rho=1000~$kg/m$^3$ is the propagation medium's density, $j_n$ are spherical  Bessel functions of the first kind, $Y_n^m=P_n^m(\cos\theta)e^{im\varphi}$ are complex spherical harmonics that involve the associated Legendre polynomials, $P_n^m$. $\omega=2 \pi f$ is the angular frequency, $k=\omega/c$ is the wavenumber, and $c=1485~$m/s is the speed of sound in water at $20^\circ$ Celsius. The beam shape coefficients $A_n^m$ allow to model the propagation properties (amplitude and phase) of the incident ultrasonic field in the spherical basis. They can be calculated semi-analytically for focused vortex fields obtained with a concave transducer geometry \cite{baresch_spherical_2013}. Applied to our case, Supplementary Fig. \ref{fig_supp_4}, shows the very good agreement between the propagation model and measurements of the incident beam magnitude, $|\tilde p|$, and indicate that sufficiently good experimental conditions are met to proceed further with the radiation force modeling. Though it is expected that sound propagates slightly faster in Carbopol hydrogels compared to pure water \cite{lidon_measurement_2017}, our results also show this increase has negligible effects on the beam propagation characteristics.

The scattered field is expanded in the same basis, leading to the scattered velocity potential, 
\begin{align}
 \phi_s&= \phi_0\sum_{n = 0}^{\infty}\sum_{m = -n}^{n}R_nA_n^m h_n^{(1)}(kr) Y_n^m(\theta,\varphi),
\label{eq:phis}
\end{align}
where $h_n^{(1)}$ are spherical Hankel functions of the first kind, and $R_n$ are the scattering coefficients for an elastic sphere embedded in an elastic medium \cite{ilinskii_acoustic_2018}, and depend on its density, $\rho_b$ shear $G_b$ and bulk $K_b$ modulii. The case of a microbubble can easily be considered by taking $G_b \rightarrow 0$. 

\begin{figure}
\centering
\includegraphics[width=0.5\linewidth]{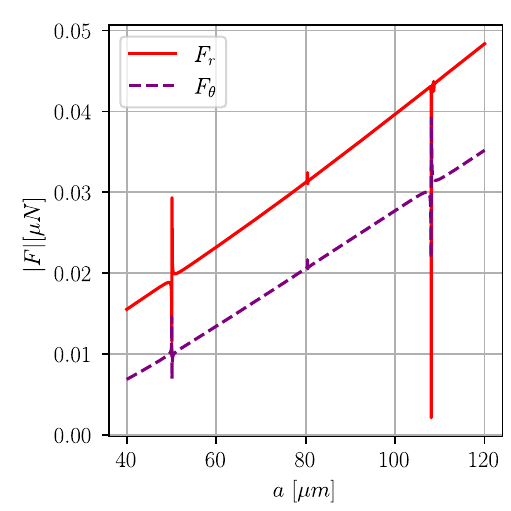}
\caption{Comparison of the modeled pressure magnitude (plain line) with hydrophone measurements (red dots) obtained for the incident vortex beam ($\ell=1, f=2.25$~MHz) in the focal plane. The hydrogel concentration, $C$, has no observable effect on the force magnitude for the soft gels we used ($G<100$~ Pa).}
\label{fig_supp_7}
\end{figure}

In a final step, following the method described in Ref.\cite{baresch_three-dimensional_2013}, the force (nonlinear) is calculated from the total field (linear) for any location of the bubble relative to the incident vortex beam. The expressions for the force components in the soft solid case \cite{ilinskii_acoustic_2018} are found to be identical to those provided in \cite{baresch_three-dimensional_2013}, the only difference residing in the computation of the scattering coefficients, $R_n$. We also assume that the bubble undergoes linear, undamped, oscillations, which is a good assumption in our experiments since the bubbles are driven far from resonance with a moderate acoustic pressure field ($p<110$~kPa), similar to recent experiments performed in pure water reporting linear oscillations \cite{baresch_acoustic_2020}. Supplementary Fig.~\ref{fig_supp_7} shows the predicted radial and azimuthal force components as a function of bubble size, $a$, in the range explored in the experiments. The general trend is that the force behaves linearly with the bubble radius in this regime. The radial force also results to be slightly larger than the azimuthal component. The curves also show marked resonance peaks at particular bubble radii. While they can result in a striking amplification of the measured bubble displacement (data not shown), we did not exploit this resonant response further in this work.     It is also worth mentioning that we did not control the exact gas mixture making the microbubbles generated by electrolysis of the hydrogel. The exact content of hydrogen and oxygen ultimately sets the value of $\rho_b$ and $K_b$ and affects the computed force. We numerically investigated the two limiting cases of pure bubbles of hydrogen and oxygen respectively (data not shown). Surprisingly, we find that the exact value of the force is only affected through the position and amplitude of the resonance peaks. We conclude that, out of the resonant regime, the force magnitude is possibly due to a variation of the scattering cross section that is only sensitive to the bubble size for gas mixtures for with the density is order of magnitudes lower compared to that of the surrounding medium.

\clearpage

\section{Shear modulus of the Carbopol hydrogels obtained with bulk rheometry}
\begin{figure}[h!]
\centering
\includegraphics[width=0.5\linewidth]{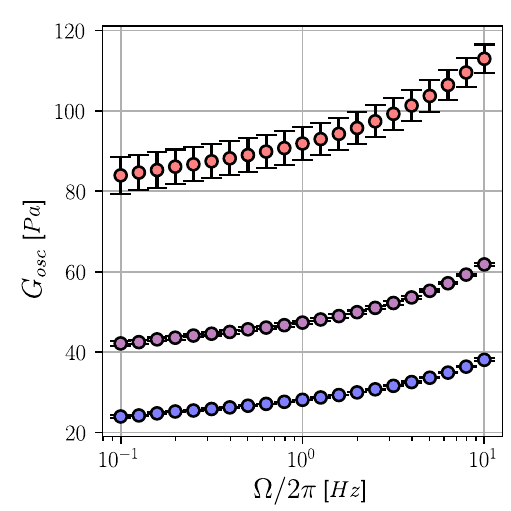}
\caption{Shear modulus obtained with oscillatory shear rheology measurements for Carbopol hydrogels of concentration $C=0.1, 0.3$ and 1\%~wt.}
\label{fig_supp_5}
\end{figure}

The elastic shear modulus of the hydrogels is characterized performing oscillatory bulk measurements using a rotational shear rheometer. The value of yield-stress of the different Carbopol hydrogels is first determined with a flow curve \cite{lidon2019mesoscale, saint2020acoustic}. The oscillatory shear stress is then set to not exceed this value, in order to obtain reliable measurements of the shear  modulus, $G_{osc}$, in the elastic regime where $\sigma = G_{osc}\gamma$. Supplementary Fig.~\ref{fig_supp_5} shows the results obtained for three different hydrogels made with $C=0.1, 0.3$ and 1.0\% wt Carbopol concentrations. The low-frequency value $G_{osc}(\Omega \rightarrow 0$ is used for comparison with the local values obtained with acoustical tweezers.

\section{Reversible and irreversible displacement of the microbubble}

\begin{figure}
\centering
\includegraphics[width=1.0\linewidth]{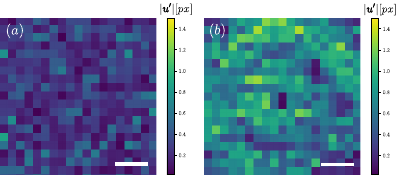}
\caption{Residual microbubble displacement maps $\mathbf{u'}$ in a Carbopol hydrogel ($C=0.1$\%~wt). (a) Low magnitude and (b) high magnitude forcing. Scale bars represent $300~\mu$m.}
\label{fig_supp_6}
\end{figure}

We compute the residual displacement, $\mathbf{u'(\mathbf{r_0})} = \mathbf{r_0} - \mathbf{r_0'}$, where $\mathbf{r_0}$ and $\mathbf{r_0'}$ are respectively the centroid position obtained before the acoustic excitation is turned on, and after it is switched off and the bubble allowed to relax to a new position. Supplementary Fig.~\ref{fig_supp_6}(a) shows that for a sufficiently low-amplitude radiation force, $ |\mathbf{u'}|$ remains small  and fluctuates around the typical position detection noise ($\sim 0.4$ pixels $\sim 0.2~\mu$m), suggesting the hydrogel followed a reversible deformation. Conversely, when the radiation force is too strong, the residual displacement $ |\mathbf{u'}|$ irreversibly exceeds the position detection noise and clearly indicates the positions at which the hydrogel yielded. Therefore, we systematically use in our experiments residual displacement maps to test that the linear elasticity hypothesis holds before extracing local measurements of $G$.        

\clearpage


%

\end{document}